\documentclass[superscriptaddress,nofootinbib,preprint]{revtex4-1}

\usepackage{hyperref}
\usepackage{bbm}
\usepackage{amsfonts}
\usepackage{mathrsfs}

\usepackage{amsmath,amssymb}
\usepackage{epsfig,amsmath,amssymb,verbatim,mathrsfs,array,layout,textcomp,amssymb,latexsym, slashed,graphicx}

\usepackage{epsfig}
\usepackage{graphicx}               
\usepackage{url}
\usepackage{hyperref}
\usepackage{float}
\usepackage{color}
\usepackage{multirow}
\makeatletter
\def\simgt{\mathrel{\lower2.5pt\vbox{\lineskip=0pt\baselineskip=0pt
           \hbox{$>$}\hbox{$\sim$}}}}
\def\simlt{\mathrel{\lower2.5pt\vbox{\lineskip=0pt\baselineskip=0pt
           \hbox{$<$}\hbox{$\sim$}}}}
\makeatother

\def\mysection#1{{\bf #1.} }

\newcommand{\be}{\begin{equation}}
\newcommand{\ee}{\end{equation}}
\newcommand{\bea}{\begin{eqnarray}}
\newcommand{\eea}{\end{eqnarray}}
\newcommand{\beq}{\begin{eqnarray}}
\newcommand{\eeq}{\end{eqnarray}}

\def\lsim{\mathrel{\rlap{\lower4pt\hbox{\hskip1pt$\sim$}}
     \raise1pt\hbox{$<$}}}         
\def\gsim{\mathrel{\rlap{\lower4pt\hbox{\hskip1pt$\sim$}}
     \raise1pt\hbox{$>$}}}         

\newcommand{\rhovac}{\sigma}

\begin{document}

\thispagestyle{empty}

\title{Spacetime Fluctuations in AdS/CFT}
\author{Erik Verlinde}
\affiliation{Institute of Physics, University of Amsterdam, Amsterdam, The Netherlands}
\author{Kathryn M. Zurek}
\affiliation{Walter Burke Institute for Theoretical Physics, California Institute of Technology, Pasadena, CA USA }

\begin{abstract}
\noindent 

We compute fluctuations in the modular energy of the vacuum associated with a Rindler-wedge in AdS spacetime in the context of AdS/CFT. We discuss the possible effect of these energy fluctuations on the spacetime geometry, and on the traversal time of a light beam propagating from the boundary to the bulk and back.

\end{abstract}

\maketitle

\tableofcontents

\newpage
\section{Introduction}
\label{sec:intro}

  In a recent paper \cite{Verlinde:2019xfb} we proposed that, generically, a theory of quantum gravity motivated by the holographic principle leads to fluctuations in the spacetime geometry that are measurable at macroscopic distances. Our reasoning was based on a number of steps. As a first step we argued that, when one considers a finite region of spacetime, the energy contained in that region, even in the vacuum, fluctuates. Specifically, we considered the fluctuations in a suitably defined energy, the so-called ``modular Hamiltonian,'' associated to a causal diamond, which describes the domain of causal dependence of a given (finite) spatial region.

 The boundary of the causal diamond is locally described by a Rindler horizon, and thus carries a non-zero temperature and entropy, both of which are due to the quantum entanglement between the interior and exterior of the diamond.   A crucial ingredient in the argumentation of \cite{Verlinde:2019xfb} was that as a result the boundary of the causal diamond can be treated as a generalization of a black hole event horizon,  in the sense that it obeys the same laws of (black hole) thermodynamics. From this assumption we showed that the size of the fluctuations in the modular Hamiltonian is directly determined by the amount of entanglement carried by the microscopic degrees of freedom associated with the diamond.
  
 The second step of our reasoning was to argue that these energy fluctuations ``gravitate,'' meaning that they induce fluctuations in the spacetime geometry. In particular, it was shown that, due to the gravitational back-reaction,  the modular energy fluctuations induce fluctuations in the location of the horizon at the edge of the causal diamond.  The third step was to demonstrate that these fluctuations give rise to observable fluctuations in the phase of light between two mirrors of an interferometer that stretches from the center to the edge of the diamond.  

The goal of this paper is to provide further evidence for the arguments and results presented in our work \cite{Verlinde:2019xfb} by placing and repeating our calculation in the context of AdS/CFT.  In the setting of AdS/CFT the microscopic description is well understood, which means that the various steps in our reasoning can be explicitly verified without uncertainty about the underlying microscopics. This allows us to make use of the many results obtained in the extensive literature on this subject to verify the central arguments  and explicitly check the conclusions of our previous work.

\newcommand{\vac}{{\rm vac}}
\subsection{Energy fluctuations in the AdS/CFT vacuum state}

Empty AdS space corresponds to the ground state of the microscopic quantum theory, and is dual to the conformally invariant ground state $|\vac\rangle$ of the CFT. Conformal transformations of the CFT are in one-to-one correspondence with isometries of the AdS geometry.  The associated conserved charges $H_\xi$ are given in terms of the stress energy tensor $T_{ab}^{{}^{{}_{CFT}}}$ of the boundary CFT by 
\beq  
\label{Hxi}
H_\xi = \int \! T_{ab}^{{}^{{}_{CFT}}}  \xi^a  dV^b,
\eeq 
where the infinitesimal volume element for the boundary CFT is denoted by $dV^b$, and $\xi^a$ represents the conformal Killing vector on the boundary. While $H_\xi$ is defined in terms of the boundary CFT, it also represents the conserved charge associated with the corresponding isometry of the bulk AdS-spacetime.

All the charges $H_\xi$ annihilate the vacuum state $|\vac\rangle$, which means that $H_\xi {}_{}|\vac\rangle~=0$. 
This fact implies that the (generalized) Hamiltonians $H_\xi$ do not exhibit any vacuum fluctuations, since the matrix elements of $H_\xi^2$ are identically zero: 
\beq
\langle \vac |{}_{}H^2_{\xi}\,|\vac\rangle =0. 
\eeq 
This does not mean, however, that there are no (local) energy fluctuations in the vacuum.  Indeed, the stress tensor itself has a non-trivial two point function, 
\beq
\langle \vac |{}_{}T_{ab}^{{}^{{}_{CFT}}}{}_{}\!(x) T_{cd}^{{}^{{}_{CFT}}}\!(y){}_{}|\vac\rangle \neq 0,
\eeq
and thus fluctuates.  The charges $H_\xi$ are defined by integrating the currents \nolinebreak$\xi^a T^{{}^{{}_{CFT}}}_{ab}$ over the entire boundary. If one would integrate over only part of the boundary, then the resulting quantities would no longer annihilate the vacuum, and will in general fluctuate.   

In this paper we will consider the special  Killing vector fields $\xi_K$ that on the boundary generate conformal transformations that leave invariant a spherical co-dimension one surface and in the bulk correspond to boosts that leave invariant a Rindler horizon. These vector fields naturally split the bulk as well as the boundary in to two parts, namely the two regions that are separated by the Rindler horizon in the bulk or a spherical co-dimension one surface on the boundary. 

Let us denote the part of the boundary inside the spherical surface by $B$ and its complement by $\overline{B}$.  The conserved charge $H_{\xi_K}$ can thus be decomposed as
$ H_{\xi_K} = K-\overline{K}$
where the quantity $K$ is defined as 
\beq
\label{defK}
K= \int_{B} T^{{}^{{}_{CFT}}}_{ab}\xi_K^a dB^b,
\eeq
and $\overline{K}$ is given by an identical expression for $\ \overline{B}$. 
Here the integral is over the spherical volume $B$, such that $dB^b$ is normal to this volume and hence points in the time direction.
We will refer to the operator $K$ as the modular energy or modular Hamiltonian. As an operator it does not annihilate the vacuum state: $K|\vac\rangle \neq 0$ and hence it can fluctuate. 
In this paper we will explicitly compute these fluctuations using the dictionary of the AdS/CFT correspondence and indeed show that the result is non-zero,
\beq
\bigl\langle K^2 \bigr\rangle - \bigl\langle  K\bigr\rangle^2 \neq 0.
\eeq
It should again be emphasized that the non-zero value of these fluctuations are a consequence of the fact that we are considering a finite region of the AdS-spacetime. 

\subsection{Summary of the main results}

The modular Hamiltonian $K$ can, instead of by Eq.~(\ref{defK}), be alternatively defined in terms the density matrix $\rhovac$ describing the mixed quantum state of the CFT inside the region $B$.   Schematically, $\rhovac$ is obtained by tracing over the Hilbert space ${\cal H}_{\overline{B}}$ for the outside region $\overline{B}$,
\beq
\rhovac ={\rm tr}_{{}_{\mbox{\footnotesize${\cal H}_{\overline{B}}$}}}\Bigl(\,|vac\rangle\, \langle vac|\,\Bigr ),
\eeq
and in this case it can be shown that the density matrix takes a thermal form \cite{Casini:2011kv} in terms of the modular Hamiltonian, 
\beq
\label{defmodular}
\rhovac ={ e^{-K}\over Z}\qquad \mbox{with}\qquad Z= {\rm tr}( e^{-K}).
\eeq
 Note that the relation (\ref{defmodular}) defines $K$ only up to a constant, since one can shift $K\to K+ c$ and $Z\to e^{-c} Z$ without changing the state $\rhovac$. We will use this freedom to identify the modular Hamiltonian with
\beq
\label{JLMS-gauge}
K \equiv -\log\rhovac . 
\eeq 
With this choice the partition function $Z$ in the vacuum state $\sigma$ is simply equal to one.

The state $\rhovac$ represents a thermal state and hence it gives rise to thermal fluctuations.  The first goal of this paper is to compute the thermal fluctuations in the expectation value of the operator $K$. This will be the starting point for determining the spacetime fluctuations. 
Normally, following the standard rules of quantum statistical physics, one computes the thermal fluctuations in the energy by computing the free energy
\beq
F_\beta = -{1\over \beta} \log {\rm tr} \left( e^{-\beta K}\right)
\eeq
as a function of the inverse temperature $\beta$, and then takes the second derivative
\beq
\bigl\langle K^2\bigr\rangle- \bigl\langle K\bigr\rangle^2= -{\partial^2 \over \partial \beta^2}\Bigl(\beta F_\beta\Bigr)_{|\beta=1}.
\eeq 
In the AdS/CFT context, however, it is not easy to compute $F_\beta$ for arbitrary values of $\beta$. Instead, we will compute the free energy $F_n$ only for integer values of $\beta =n$ using the replica method. Specifically, we will use the methods developed in the references \cite{Lewkowycz:2013nqa, Hung:2011nu, Dong:2016fnf}.

First, it is clear from Eq.~(\ref{JLMS-gauge}) that the expectation value of the modular Hamiltonian $K$ is equal to the entanglement entropy, which  in a CFT that is dual to Einstein gravity can be computed using the Ryu-Takayanagi formula \cite{Ryu:2006bv,Hubeny:2007xt}
\beq
\bigl \langle K\bigr\rangle = {A(\Sigma)\over 4G}.
\eeq
Here $A(\Sigma)$ denotes the area of the extremal surface $\Sigma$ in AdS that is anchored on the boundary on the edge of the region $B$. In the present context $\Sigma$ also represents the Rindler horizon. 
A central result of this paper will be that the fluctuations in the modular Hamiltonian within AdS/CFT are given by
\beq
\bigl\langle K^2\bigr\rangle- \bigl\langle K\bigr\rangle^2 = {A(\Sigma)\over 4G}.
\eeq
This results holds in arbitrary dimensions for any strongly coupled CFT with a large $N$ gravitational dual that is described by the Einstein action. We will subsequently study the implications of these modular energy fluctuations for the AdS-Rindler horizon and for the travel times of photons that probe the spacetime geometry near the horizon.

\section{The geometric setup}

 Consider a light signal that is first emitted from the boundary and then is reflected by a ``mirror" at some location in the bulk, and is finally again received at a later moment on the boundary. The points of emission and reception on the boundary define a causal diamond that consists of all points that are in the causal future of the point of emission and the causal past of the point of reception.
The past lightcone of the reception point intersects that future lightcone of the emission point along a co-dimension two surface, that represents a Ryu-Takayanagi (RT) surface $\Sigma$ in the bulk. The causal diamond describes a wedge of Rindler-AdS space, with the Killing horizon located at its boundary.   The light signal travels along this horizon and reflects at the point on the RT surface $\Sigma$. This situation is depicted in Fig.~1.   Note that our geometric set-up is rather similar to some works seeking to derive the linearized Einstein equations from entanglement entropy \cite{Faulkner:2013ica,Jacobson:2015hqa}.

\begin{figure}[btp]
\begin{center}
\includegraphics[scale=0.51]{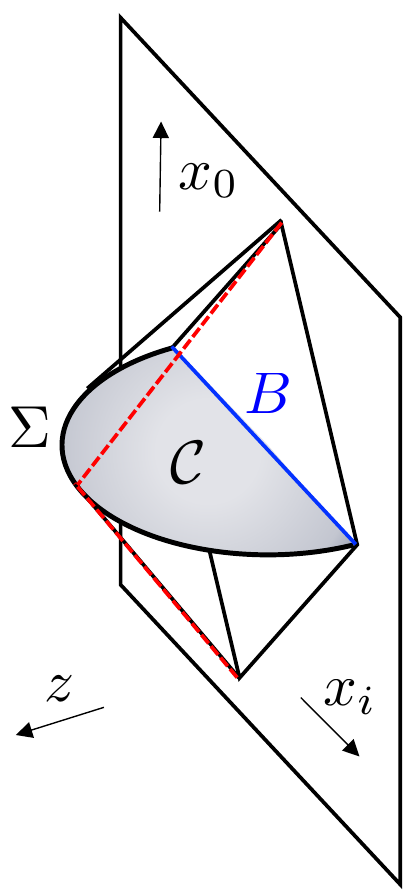}
\label{fig:causaldiamond}
\vspace{-0.5cm}
\caption{Depiction of the causal diamond in AdS space anchored at the boundary. The red dotted line is the light signal from the boundary to a point in the bulk. }
\end{center}
\end{figure}

\subsection{From Poincar\'{e} to AdS-Rindler coordinates}

On the boundary we are thus dealing with a spherically symmetric causal diamond, whose causal domain extends in the bulk and describes one wedge of Rindler-AdS space. The Rindler horizon coincides with the RT surface $\Sigma$, which is anchored on the edge of the causal diamond on the boundary.
We will work in two coordinate systems, one that describes the $(d+1)$-dimensional AdS geometry near the boundary in terms of the familiar Poincar\'{e} coordinates:
\beq
\label{Poincare}
ds^2 =L^2\, { dz^2 + dx_i^2-dx_0^2\over z^2} \qquad \quad \mbox{with} \qquad i= 1,\ldots ,d-1. 
\eeq
The finite spherical region on the boundary is described by $x_i^2\leq R^2$. The shaded region in the bulk represents the spatial slice inside the causal diamond and is given by $z^2 + x_i^2 \leq R^2 $. 
The full interior of the causal diamond is described by
\beq
\label{causaldomain}
R^2-z^2 - x_i ^2 +x_0^2\ \geq  2R |x_0|.
\eeq
It turns out that, while the region of interest is easily described in these coordinates, the calculation of the fluctuations in the modular Hamiltonian and the possible backreaction effects are more easily studied in a different coordinate system that is better adapted to the geometry of the causal diamond. 

For this reason we introduce a second coordinate system that describes the interior of the causal diamond and puts it in the form of Rindler-AdS space (see for example Ref.~\cite{Casini:2011kv} for more details).   The associated metric is\\[-4mm]
\beq
ds^2 = -\left({r^2\over L^2}-1\right)dt^2 + \left({r^2\over L^2}-1\right)^{-1} \!dr^2 + r^2 d\Sigma_{d-1}^2,\\[-8mm]
\eeq
where\\[-12mm] 
 \beq
 \label{hyperbolic}
d\Sigma^2_{d-1} = d\chi^2 +\sinh^2\!{\chi}\,d\Omega_{d-2}^2\\[-8mm]
\eeq 
describes the geometry of a hyperbolic plane.
  The radial coordinate obeys $r\geq L$, so that the coordinate $t$ remains timelike. The boundary value $r=L$ corresponds to the location of the Rindler horizon. The relationship between the two coordinate systems is\footnote{This coordinate transformation can be derived by making use of the embedding equations for AdS and tracing through the conformal mapping used in Ref.~\cite{Casini:2011kv} to go from Rindler space to the causal diamond. }  
\beq
{R^2-z^2-x_i^2+x_0^2\over 2R z}= \left({r^2\over L^2}-1\right)^{1\over 2} \cosh{t\over L},
\qquad\qquad
{x_0\over z}= \left({r^2\over L^2}-1\right)^{1\over 2} \sinh{t\over L}. \label{coordinatechange}
\eeq 
The spatial $x_i$ coordinates are replaced by the coordinate $\chi$ and the solid angles $\Omega$ by writing the norm $|x^i|$ as\\[-12mm]
\beq
\label{hyperboliccoordinates}
|x_i| = R\tanh \chi.
\eeq
One easily verifies that new coordinates fully cover the domain (\ref{causaldomain}) and that the $r=L$ locus coincides with the boundary of this causal domain.  The new $(r,t, \chi, \Omega)$-coordinate system thus describes only the interior of the causal diamond.


\subsection{The regularized horizon area of AdS-Rindler and topological black holes}

A central quantity in our calculations is the area of the AdS-Rindler horizon, which through the result of Ryu and Takayanagi measures the entanglement entropy between inside and outside of the causal diamond for the boundary CFT,
\beq
\label{RT}
S= {A(\Sigma)\over 4G}.
\eeq
The surface $\Sigma$ is located at radius $r=L$. This means that we can write its area as
\beq
\label{areadef}
A(\Sigma) = L^{d-1} V_{d-1},
\eeq
where $V_{d-1}$ denotes the dimensionless `volume' of the $(d-1)$-dimensional hyperbolic plane described by the metric Eq.~(\ref{hyperbolic}). The hyperbolic volume $V_{d-1}$ is formally infinite, and hence needs to be regularized.
We will therefore introduce a cut-off by restricting the range of the coordinate 
$
\chi\leq \chi_c. 
$ 
This leads to the following regularized expression\\[-6mm]
\beq
V_{d-1} = \Omega_{d-2} \int_0^{\chi_c} \!\!\! d\chi \,  \sinh^{d-2} \chi,
\label{regularizedvolume}
\eeq
where $\Omega_{d-2}$ is the area of a unit $(d\!-\!2)$-sphere.  


A further important reason for introducing this regularization is that it allows us to consider geometries that generalize AdS-Rindler space by 
allowing the (modular) energy $K$ to differ from its vacuum value. These 
geometries are called topological black holes and are described by the 
metric
\beq
\label{topBH}
ds^2 = -f(r)dt^2 + {dr^2\over f(r)} + r^2 d\Sigma_{d-1}^2,
\eeq
where the blackening factor $f(r)$ takes the form
\beq
\label{fofr}
f(r) = {r^2\over L^2}-1- 
2 \Phi{ \,L^{d-2} \over r^{d-2}}, 
\eeq
obtained by solving the Einstein equations.  The dimensionless quantity $\Phi $ is an integration constant, which physically represents the value of a Newtonian potential on the original location $r=L$ of the horizon.  $\Phi$ can be written in terms of a mass $M$, which we will show below is the mass measured by the observer who uses $t$ as their time coordinate.  In this case, we will have\\[-8mm]
\beq
\label{PhitoM}
\Phi = {8\pi GM\over (d\!-\!1)V_{d-1}L^{d-2}},
\eeq
where again $V_{d-1}$ is the regularized hyperbolic volume of the horizon. A similar regularization is required to define the mass $M$, which without any regularization would be infinite.  The ratio $M/V_{d-1}$ is independent of the regularization and may be interpreted as the tension of a brane that lives at the location of the horizon. In fact, in the following we will make use of these geometries, with precisely this brane interpretation, to calculate the fluctuations in the values of the modular Hamiltonian.

\section{Fluctuations in the modular Hamiltonian}

Our goal is to compute the fluctuations in the modular Hamiltonian $K$ in the vacuum state $\rhovac$. These fluctuations are described by the two point function,
\beq
\bigl\langle (K  - \langle K \rangle)^2\bigr\rangle = {\rm tr}\bigl(\rhovac (K-\langle K\rangle)^2\bigr) \qquad \mbox{with} \qquad \langle K\rangle = {\rm tr}\bigl(\rhovac K\bigr).
\eeq
But first we will compute its expectation value $\langle K\rangle$ and review the relationship between the boundary and bulk modular Hamiltonian. 

\subsection{The modular Hamiltonian: bulk versus boundary} 

We have defined the modular Hamiltonian  $K$ by the relation (\ref{JLMS-gauge}) in terms of the density matrix $\rhovac$. This definition implies that the vacuum expectation $\langle K\rangle $ is non-zero. Indeed, one can easily show that $\langle K\rangle $ is equal to the entanglement entropy $S$,
\beq
\bigl\langle K\bigr\rangle = {\rm tr}\bigl(\rhovac K \bigr)  = -{\rm tr}\bigl(\rhovac \log\rhovac \bigr) = S.
\eeq
Together with the RT formula we may thus conclude that for a CFT with an AdS dual described by Einstein gravity that
\beq
\bigl\langle K\bigr\rangle = {A(\Sigma)\over 4G}.
\eeq
As a next step we introduce a new operator $\Delta K$ that is defined as the deviation of the modular Hamiltonian $K$ from its vacuum expectation value,
\beq
\Delta K \equiv K-\langle K\rangle \cdot {\mathbf 1}. 
\eeq
 This operator is well behaved in the sense that it has no divergences, and its vacuum expectation value is by definition equal to zero. 
 With the above identifications, the definition of $\Delta K$ can be 
alternatively written as 
\beq
K = {A(\Sigma)\over 4G}  \cdot {\mathbf 1} + \Delta K.
\eeq 
A noteworthy property of the operator $\Delta K$ is that its expectation value, when computed in a state $\rho$ that differs from the vacuum state $\rhovac$, is equal to the relative entropy. One has
\beq
{\rm tr}\left(\rho \Delta K\right) = -{\rm tr}\left(\rho \log \rhovac\right) +{\rm tr}\left(\rhovac \log \rhovac\right) \equiv S_{rel}(\rho|\rhovac). 
\eeq
This result is familiar from Ref.~\cite{Jafferis:2015del}, and reflects the fact that the boundary relative entropy equals the bulk relative entropy. 
Furthermore, according to \cite{Jafferis:2015del}, the quantity $\Delta K$ represents the canonical energy of the bulk theory, which takes the form of the bulk modular Hamiltonian\\[-10mm]
\beq
\Delta K = \int_{\cal C} \xi^\mu_K T^{bulk}_{\mu\nu} d{\cal C}^\nu,  
\eeq
where ${\cal C}^\nu$ is the volume denoted by the shaded area in Fig.~1. 
It is clear from this expression that $\Delta K$ sources the gravitational field in the bulk. In other words, it gravitates. 
Interestingly, it is also the same quantity that appears in the two-point function for the vacuum fluctuations of the modular Hamiltonian: 
\beq
\label{fluctuate}
\bigl\langle \Delta K^2 \bigr\rangle = 
\bigl\langle (K  - \langle K \rangle)^2\bigr\rangle. 
\eeq
Our goal is to determine these fluctuations by applying the AdS/CFT dictionary. As as first step we note that, by inserting the definition Eq.~(\ref{JLMS-gauge}) of $K$, one can write the right-hand-side of Eq.~(\ref{fluctuate}) explicitly in terms of the state $\sigma$ as 
\beq
\label{DeltaKfromsigma}
\bigl\langle \Delta K^2\bigr\rangle = {\rm tr}\bigl(\rhovac (\log\rhovac)^2\bigr)- \Bigl({\rm tr}\bigl(\rhovac \log\rhovac\bigr)\Bigr)^2.
\eeq
This equation manifestly shows that all the necessary information to compute $\langle \Delta K^2\rangle$ is contained in the state $\sigma$ itself. We will follow this path in the next subsection, where we apply the gravitational version of the replica method in which only the state $\sigma$ is used as input.

\subsection{The gravitational replica method}

We will calculate the expectation value and the fluctuations in the modular Hamiltonian $K$ by making use of a gravitational version of the replica method developed in \cite{Lewkowycz:2013nqa, Hung:2011nu, Dong:2016fnf}.  Instead of introducing a finite inverse temperature $\beta$, we define free energies $F_n$, 
\beq
\label{defFn}
F_n  = -{1\over n}\log\Bigl({\rm tr}\bigl(\rhovac^n\bigr) \Bigr),
\eeq
where $n$ is first chosen to take integer values\footnote{The free energies $F_n$ are closely related to the R\'{e}nyi entropies $ S^{Renyi}_n = {1\over 1-n} {\rm tr}\log \rhovac^n $. }.  One can think of the free energies $F_n$ as being associated to thermal states $\rho_n$ with inverse temperature equal to $n$ times that of the vacuum state $\rhovac$,
\beq
\label{rhon}
\rho_n = {\sigma^n\over {\rm tr} \bigl(\sigma^n\bigr)} = {e^{-nK}\over Z_n}
\qquad
\mbox{with} 
\qquad Z_n ={\rm tr}(e^{-nK}) = e^{-nF_n}.
\eeq 
The partition function $Z_n$ is related to the free energy $F_n$ in the standard fashion. 

 The purpose of this section is to calculate the free energies $F_n$ in AdS/CFT. Fortunately, this calculation was already performed in Refs.~\cite{Hung:2011nu,Dong:2016fnf}
by extending the method of Ref.~\cite{Lewkowycz:2013nqa}. We will review these computations here: it will produce an expression for $F_n$ for all integer values of $n$. Afterwards we will formally treat $n$ as a continuum variable, and view $F_n$ as an analytic function of $n$. This allows us to subsequently take its derivatives with respect to $n$. As we will explain below, the expectation value and fluctuations in $K$ can be expressed in terms of these derivatives.

The gravitational replica method developed in Refs.~\cite{Lewkowycz:2013nqa,Hung:2011nu,Dong:2016fnf} amounts to considering the path integral over $n$-fold covers of an asymptotically AdS-Rindler spacetime. The gravitational path integral can be evaluated in the saddle point approximation, where the dominant saddle point represents a ``cosmic brane'' solution that is branched over the locus of the Rindler horizon \cite{Dong:2016fnf}. In this way one finds that the free energy $F_n$ can be expressed in terms of the classical gravitational action $I_n$ for the cosmic brane, 
\beq
nF_n = I_n - nI_1.
\eeq 
By requiring that the saddle point solution remains regular at the branch point, one learns that the underlying geometry of the base space, which is covered $n$-times by the full geometry, has an angle deficit equal to\\[-12mm]
\beq
\label{angledeficit}
\Delta\phi_n = 2\pi \left(1-{1\over n}\right).
\eeq     
The Lorentzian geometry of a cosmic brane with angle deficit (\ref{angledeficit}) takes the form of the topological black hole metric Eq.~(\ref{topBH}),
where the blackening function 
is given by 
\beq
\label{fn}
f_n(r) = {r^2\over L^2}-1 - 2\Phi_n {  L^{d-2}\over r^{d-2}}.
\eeq
As explained in the discussion around Eq.~(\ref{PhitoM}), the quantity $\Phi_n$ is related to the tension (or energy density) of the cosmic brane. In this situation the required cosmic branes turn out to have negative tension.  

The horizon of this geometry is located at $ r = r_n$, where $r_n$ is the solution to 
\beq
\label{horizon}
f_n(r_n) =0.
\eeq 
We will denote the co-dimension-two surface given by the bifurcate  horizon of the cosmic brane by $\Sigma_n$. For $n=1$ it coincides with the original Rindler-AdS horizon $\Sigma$. 
The inverse temperature $T_n$ of the cosmic brane solution (or topological black hole) is given by
\beq
\label{inversetemp}
T_n = {f_n'(r_n)\over 4\pi}. 
\eeq
Since there is no cosmic brane for the case $n=1$, we know that $\Phi_1=0$: in this case one finds $T_1 = 1/2\pi L$. By imposing that the angle deficit of the cosmic brane takes the value $\Delta \phi_n$  given in Eq.~(\ref{angledeficit}), one deduces that the temperature $T_n$ obeys
\beq
\label{betan}
T_n = {1\over 2\pi L n}. 
\eeq
This equation, together with equation (\ref{horizon}), forms a coupled set of equations from which both the radius $r_n$ as well as the value of $\Phi_n$ can be determined. These equations are most conveniently written in terms of the dimensionless numbers $x_n= r_n/L,$ where normalization is chosen so that $x_1=1$. 
 By requiring that the inverse temperature $T_n$ defined in Eq.~(\ref{inversetemp}) gives the result Eq.~(\ref{betan}), one learns that  the variables $x_n$ are given by the positive solutions to the quadratic equation,
\beq
\label{quadratic}
d\, x^2_n - {2\over n} x_n- (d-2) = 0 
\qquad \mbox{with}\qquad x_n=r_n/L. 
\eeq
It is a simple exercise to obtain the solution to this equation and write an explicit expression for $x_n$ in terms of $n$ and $d$. Since the result will not be particularly useful for the following discussion, we leave this exercise to the reader. Instead we will use $x_n$ itself to express the various other physical quantities. For instance, one can compute the value of $\Phi_n$ by imposing the condition that $f(r_n)=0$. This leads to the result 
\beq
\label{Phin}
\Phi_n = {1\over 2}\left(x_n^2-1\right) x^{d-2}_n.
\eeq
This relation can be used to subsequently eliminate $\Phi_n$ from the remaining equations, so that all variables are expressed in terms of $x_n$. 

It should be noted that all positive solutions to Eq.~(\ref{quadratic}) are less than one and only equal to one when $n=1$. Hence, $x_n<1$ for $n>1$. This implies that $\Phi_n$ is negative, which also means that the ``cosmic branes'' described by the topological black hole geometries Eqs.~(\ref{topBH}) and (\ref{fn}) have a negative tension and hence negative mass $M$.  

Our next goal is to calculate the free energies $F_n$. In principle, this can be done by evaluating the classical action of the cosmic brane solutions. It appears, however, that there is a more direct and simple approach. Instead of directly calculating the Euclidean action $I_n$ of the cosmic brane solutions, we follow the method of Ref.~\cite{Hung:2011nu}. By applying the familiar thermodynamic  relation for the free energy $F(T)= E-TS$ together with the first law $dE=TdS$ one can derive the following integral expression for the free energy\footnote{The dimensionless free energies $F_n$ are related to the conventional free energy $F(T)$ via $nF_n= F(T_n)/ T_n$. }
\beq
\label{integral}
nF_n = {1\over T(x_n)}\left\lbrack -T(x)S(x) \Bigl |^{x_n}_1 +\int^{x_n}_1 \!\! T(x) dS(x)\right\rbrack, 
\eeq
where the functions 
\beq 
S(x)= {A(\Sigma)\over 4G} x^{d-1} \qquad \quad
\mbox{and}\qquad \quad
T(x) ={1\over 4\pi L} \left( d\, x - {d-2\over x}\right) 
\eeq 
represent the entropy and temperature associated with the horizon at $r/L=x$. The integral in this equation is easily evaluated, and together with the first term gives the following result for the free energy: 
\beq
\label{Fn}
F_n = {A(\Sigma)\over 4G} \left(1-{1\over 2}x^{d}_n -{1\over 2}x_n^{d-2}\right).
\eeq
With this result it now becomes straightforward to compute the fluctuations in the modular Hamiltonian $K$.

\subsection{Computation of the fluctuations in the modular Hamiltonian}

The vacuum fluctuations of the modular Hamiltonian $K$ as well as its vacuum expectation value can both be expressed as derivatives of the free energy $F_n$. First of all, the vacuum expectation value equals
\beq
\bigl \langle K\bigr\rangle =  {d \over dn}\Bigl(n F_n\Bigr)\Bigr|_{n=1}
\eeq
One easily verifies this relation with the help of Eq.~(\ref{JLMS-gauge}). In a similar way one obtains the fluctuations in $K$ by differentiating twice with respect to $n$. One has 
\beq
\bigl\langle \Delta K^2\bigr\rangle = -{d^2 \over dn^2}\Bigl(n F_n\Bigr)\Bigr|_{n=1}
\eeq
To check this last relation one inserts Eq.~(\ref{JLMS-gauge}) into the left-hand-side, or uses the definition Eq.~(\ref{defFn}) for $F_n$ on the right-hand-side.  

 To evaluate these derivatives of the free energy $F_n$ with respect to $n$, we use its expression Eq.~(\ref{Fn}) in terms of the variables $x_n$.  First let us  determine the expectation value of the modular energy $K$ in the infinite set of generalized thermal states $\rho_n$ defined in Eq.~(\ref{rhon}).  
 These expectation values are given in terms of the free energy $F_n$ by
\beq
\bigl \langle K\bigr\rangle_n \equiv {\rm tr}\bigl(K \rho_n \bigr) ={{\rm tr} \bigl( K \sigma^n\bigr)\over {\rm tr}(\sigma^n)} 
={d \over dn}\Bigl(n F_n\Bigr).
\eeq
Here we keep $n$ free. 
To compute $\langle K \rangle_n$ for general values of $n$ we need to know the derivative of $x_n$ with respect to $n$. By multiplying the quadratic relation Eq.~(\ref{quadratic}) by $n/2x_n$ and differentiating with respect to $n$ one finds: 
 \beq
\label{identity}
 n \left(d\, x_n + {d-2\over x_n}\right) {dx_n\over dn} =-d \,x_n + {d-2\over x_n}.
 \eeq 
 With the help of this relation one can now easily compute the vacuum expectation value of $K$ and its vacuum fluctuations. First of all, for the expectation value $\langle K\rangle_n$ we find
\beq
\label{DeltaKn}
\bigl \langle K\bigr\rangle_n = {A(\Sigma)\over 4 G} \Bigl(1 +{d-1\over 2}\left( x_n^2-1\right) x_n^{d-2} \Bigr).
\eeq
By taking $n=1$ one verifies that the vacuum expectation value of $K$ is, as expected, equal to the entanglement entropy 
\beq
\label{Kvac}
\bigl\langle K \bigr\rangle = {A(\Sigma)\over 4G}.
\eeq
The next step is to evaluate the fluctuations. Using the representation (\ref{Fn}) it is straightforward to show that
\beq
\bigl\langle \Delta K^2  \bigr\rangle= -{d^2\over dn^2}\Bigl( n F_n\Bigr)\Bigr|_{n=1}= -{d\over dn} \bigl \langle K\bigr\rangle_n \Bigr|_{n=1} = -{1\over n} {d\over dn} S(x_n)\Bigr|_{n=1}, 
\eeq
where in the last step we made use of the first law of thermodynamics. 
This last expression can also be directly derived by differentiating the integral
expression (\ref{integral}) twice with respect to $n$ and make use of the fact that $1/T(x_n) = 2\pi L n$.
The right hand side can be evaluated in a straightforward way and gives
\beq
{d\over dn} S(x_n)\Bigr|_{n=1} = (d-1){A(\Sigma)\over 4G}  x_n^{d-2}{dx_n\over dn}\Bigr|_{n=1} = -{A(\Sigma)\over 4G}, 
\eeq
where the derivative of $x_n$ is evaluated via the relation (\ref{identity}): $dx_n/dn|_{n=1} = -1/(d\!-\!1) $.\linebreak In this way we arrive at the final result for the fluctuations in the modular Hamiltonian\footnote{After completing our work we learned of related papers 
in a similar result was obtained, see {\it Note added.}}
\beq
\label{mainresult}
\bigl\langle \Delta K^2  \bigr\rangle = {A(\Sigma)\over 4G}. 
\eeq
This is indeed the answer that we anticipated in \cite{Verlinde:2019xfb}, and can be argued for on general grounds, since it expresses the fact that the size of the fluctuations $\sqrt{\langle \Delta K^2\rangle}$ obeys the familiar ``square root" law for a gaussian stochastic process.

\section{Backreaction due to the fluctuating modular energy}

After having established that the modular Hamiltonian fluctuates in the vacuum, we now come to the question of how these fluctuations influence the geometry. 
As we made clear, the vacuum expectation value of the modular Hamiltonian does not gravitate, since this would imply that the vacuum expectation value of the geometry would not be equal to that of the vacuum. The difference $\Delta K$ between the value of the modular Hamiltonian $K$ and its vacuum expectation value does gravitate. In particular, as we will show, a state for which the expectation value of $K$ differs from that of the vacuum, will correspond to a geometry different from empty AdS. 

Specifically we will show that a non-zero expectation value of $\Delta K$ results in a non-zero (expectation) value of the parameter $\Phi$ that we identified with the Newtonian potential. To show this fact, we will first relate $\Delta K$ and the mass $M$ that appears in Eq.~(\ref{PhitoM}) for $\Phi$. Subsequently we will argue that the same relation between $\Delta K$ and $\Phi$ also holds at the level of the fluctuations. This implies in particular that $\Phi$ fluctuates and that as a result the location of the horizon exhibits vacuum fluctuations. Our reasoning in this section is inspired by \cite{Marolf:2003bb} and will closely follow the discussion in \cite{Verlinde:2019xfb}. 

\subsection{The relation between the mass $M$ and the modular hamiltonian $K$}
\label{subsec:NewtonianFlucs}

The aim of this subsection is to show that the mass $M$, as defined by Eq.~(\ref{PhitoM}), can, as a quantum  operator, be identified with 
\beq
\label{MtoK1}
M 
= {1 \over 2\pi L} \Bigl( K -\bigl\langle K\bigr\rangle \Bigr).
\eeq
Here the  pre-factor $1/2\pi L$ represents the Hawking temperature at the Rindler horizon. 
In other words, the mass operator $M$ is, up to an overall factor given by the temperature, equal to difference of the modular Hamiltonian $K$ and its vacuum expectation value $\langle K\rangle$. A crucial ingredient in the derivation of this relation is the fact that the mass $M$ satisfies the first law of black 
hole thermodynamics
\beq
\label{dMisTdS}
d M = T d S. 
\eeq
Our proof of Eq.~(\ref{MtoK1}) amounts to showing that both sides of this equation obey the same 1st law of black hole thermodynamics.

First, let us derive that the mass $M$, as defined by Eq.~(\ref{fofr}) and Eq.~(\ref{PhitoM}), indeed  obeys the first law given in Eq.~(\ref{dMisTdS}). For this we use the fact that the location of the horizon is determined by the equation $f(r_h,M) = 0$ for all values of $M$. Hence, the variation $dr_h$ of the horizon location and the mass variation $dM$ are related by
\beq
d f(r_h,M) = f'(r_h,M) d r_h + \frac{\partial f(r_h,M)}{\partial M} dM =0.
\eeq
From the expressions in Eq.~(\ref{fofr}),~(\ref{PhitoM}) we learn that 
\beq
\frac{\partial f(r_h,M)}{\partial M} = - \frac{16 \pi G}{(d-1) V_{d-1} r_h^{d-2}}.
\eeq
It is now a simple exercise to show, with the help of Eq.~(\ref{inversetemp}) and by varying the area-entropy relation in Eq.~(\ref{RT}), that the variation $d M$ indeed satisfies the first law of black hole thermodynamics given in Eq.~(\ref{dMisTdS}). 

Our next goal is to show that the right-hand side of Eq.~(\ref{MtoK1}) obeys the same first law of black hole entropy, when $K$ is defined by Eq.~(\ref{JLMS-gauge}). We will do this at the level of expectation values in the infinite set of states $\rho_n$ defined in Eq.~(\ref{rhon}). 
For this purpose it will be useful to rewrite Eq.~(\ref{dMisTdS}) in terms of the free-energy $F(T)=M-TS$ in the familiar form\\[-8mm]
\beq
\label{dFSdT}
dF = -SdT 
\eeq
To show that this relation also holds for the right-hand side of Eq.~(\ref{MtoK1}) we consider the expectation value in the infinite set of states $\rho_n$ defined in Eq.~(\ref{rhon}). First of all, one easily verifies that the dimensionless free energies $F_n$ defined in Eq.~(\ref{defFn}) obey 
\beq
F_n \equiv  -{1\over n} \log {\rm tr} \left(\sigma^n\right) = 
\bigl\langle K\bigr\rangle_n - {1\over n} S_n,
\eeq
where $S_n= -{\rm tr}\left(\rho_n\log\rho_n\right)$
is the von Neumann entropy\footnote{Note that the entropies $S_n$ differ from the R\'{e}nyi entropies $S_n^{Renyi}$ defined in the footnote below Eq.~(\ref{defFn}).} associated with the state $\rho_n$.  By taking the differential of the first expression one derives that $F_n$ satisfies 
\beq
\label{dFdSn}
d F_n = 
S_n {dn\over \ n^2}
\eeq
which is the direct analogue of Eq.~(\ref{dFSdT}).
In these steps $1/n$ plays the role of a dimensionless temperature, and is considered to be a continuous variable.  This can be made even more explicit by rewriting this relation, using the second expression for $F_n$ in Eq.~(\ref{Fn}), in the form
\beq
d \bigl\langle K\rangle_n  = {1\over n} \,dS_n.
\eeq
This is the direct analogue of the first law in Eq.~(\ref{dMisTdS}). Now note that the physical temperature $T_n$ defined in (\ref{betan}) differs from $1/n$ by a factor $2\pi L$.  The same factor appears therefore in the relation between the mass $M$ and the modular Hamiltonian $K$. Finally, by requiring that the mass $M$, defined as an operator, has a zero vacuum expectation value, one uniquely fixes the relation between the mass and the modular Hamiltonian to take the form Eq.~(\ref{MtoK1}). 


As a final check on this identification, we note that inside the causal diamond on the boundary the modular operator $K$ acts as the generator of a conformal boost in the CFT, while in the bulk it represents a Lorentz boost that leaves the Rindler horizon invariant. The corresponding (conformal) Killing vector $\xi_K$ is related to the generator $\partial/\partial t$ of translation in AdS-Rindler time by a factor equal to the temperature associated with the Rindler horizon. Hence, we have\\[-12mm]
\beq
\xi^\mu_K \partial_\mu =2\pi L {\partial \over \partial t}. 
\eeq
This equation again implies that the factor $2\pi L$ must also appear in the relation between the mass $M$ and the quantity $K-\langle K\rangle $. 

\subsection{Fluctuations in the Newtonian potential on the Rindler horizon}
\label{subsec:NewtonianFlucs}

In this subsection we take the next step to show that the fluctuations in the modular Hamiltonian associated with the causal region also leads to metric fluctuations, specifically in the Newtonian potential $\Phi$. 
By inserting this relation into the expression Eq.~(\ref{PhitoM})  we arrive at the expression for $\Phi$,
\beq
\label{PhitoK}
 \Phi =  {\Delta K\over  (d\!-\!1)}   {4G \over A(\Sigma)}\qquad\quad \mbox{with}\quad\qquad \Delta K = K-\bigl\langle K\bigr\rangle. 
\eeq
As announced, this relation shows that $\Delta K$ gravitates.   Since $\Delta K$ has a zero vacuum expectation value, it induces no change in the expectation value of the metric, but the fact that $\Delta K$ exhibits vacuum fluctuations means that it can lead to small quantum fluctuations in the value of the Newton potential $\Phi$ on the AdS-Rindler horizon.

Just as in our discussion for the mass $M$ we will again read this equation as an operator identity relating $\Phi$ to $\Delta K$. In making this identification, we will regard the area of the Rindler horizon $A(\Sigma)$ as a scalar quantity that is fixed: it represents the size of the Hilbert space and the entanglement entropy of the vacuum state.  As such it is not a quantum operator. 

To further justify the relation Eq.~(\ref{PhitoK}) we will first verify it again at the level of expectation values in the state $\rho_n$. We want to verify that the expectation value  $\bigl\langle \Phi \bigr\rangle_n $ of $\Phi$ in the state $\rho_n$ is given by the  cassical result that we determined in Eq.~(\ref{Phin}). In other words, we want to show that we can take $\bigl\langle \Phi \bigr\rangle_n  \! \equiv \Phi_n $. 
To verify this relation we compute the expectation value of the right-hand side of Eq.~(\ref{PhitoK}). This gives
\beq 
 \bigl\langle \Phi\rangle_n =  {\bigl\langle \Delta K\bigr\rangle_{n}\over  (d\!-\!1)} {4G \over A(\Sigma)}.  
\eeq
By comparing the expressions (\ref{Phin}) and (\ref{DeltaKn}) for $\Phi_n$ and $\langle K\rangle_n$, and after subtracting the vacuum value Eq.~(\ref{Kvac}) for $\langle K\rangle$, one easily checks that the right-hand side is equal to $\Phi_n$, as required. 
This therefore suggests that Eq.~(\ref{PhitoK}) also holds at the level of operators.  

If we assume that this operator identification indeed holds, we can now use it to compute the fluctuations in the Newton potential $\Phi$ in a simple and straightforward way. We find that the value of $\Phi$ on the horizon has a non-zero two-point function in the vacuum state $\sigma$ given by\\[-10mm]
\beq 
\label{Phisq}
\bigl \langle \Phi^2\bigr \rangle  =  {\bigl\langle\Delta K^2\bigr\rangle\over  (d\!-\!1)^2} \left( {4G \over A(\Sigma)}\right)^2 ={1\over  (d\!-\!1)^2} {4G \over A(\Sigma)}, 
\eeq
where in the last step we used our result Eq.~(\ref{mainresult}) for the fluctuations in $\Delta K$. 

After having argued that the fluctuations in the modular energy $\Delta K$ also source fluctuations in the geometry, in particular in the Newtonian potential $\Phi$ on the horizon, we will now investigate the consequences of such fluctuations. 

\subsection{Horizon location and photon travel time }

Let us return to the AdS spacetime in Poincar\'{e} coordinates
\beq
ds^2 = {L^2\over z^2}\Bigl(dz^2 + dx_i^2-dx_0^2\Bigr),
\eeq
and again consider the causal domain bounded by the AdS-Rindler horizon. The horizon can be thought of as being made up of many photon trajectories that are simultaneously emitted from the boundary, and subsequently reflect off an (imaginary) mirror located on the bifurcate horizon.   Our aim in this subsection is to use this observation to show that the travel time of a real photon, emitted from the boundary and reflected off a real mirror, is directly correlated with the location of the associated Rindler horizon. 


 The coordinate transformation from the Poincar\'{e} coordinates to the Rindler coordinates given in Eq.~(\ref{coordinatechange}) implies the following relation \\[-4mm]
\beq
\label{lightcone}
\left({(R-x_0)^2- z^2-x_i^2\over 2Rz}\right)\left({(R+x_0)^2- z^2-x_i^2\over 2Rz}\right)={r^2\over L^2}-1.
\eeq\\[-4mm]
This equation expresses when we take $r=L$ that either the first or 
the second factor on the left-hand-side vanishes. The resulting equations 
describe the location of the lower and upper parts of the causal diamond 
in the bulk. The intersection of the two equations corresponds to the 
bifurcate horizon, which is hence located at $x_0=0$ and $z^2+x_i^2=R^2$. 

We now consider a photon trajectory which travels along  $x_i=0$ and reflects off a mirror located on the bifurcate horizon at $z=R$.  We can describe its path by disregarding the $x_i$ coordinates in Eq.~(\ref{lightcone}). In this way one finds that the photon is emitted and received at the boundary at $x_0=-R$ and $x_0=R$.  This equations hold when the boundary is chosen to be located at $z=0$. However, to regulate the horizon area and to obtain a finite induced metric on the boundary, we will now introduce a small cut-off $z_c$ by restricting the range of the $z$-coordinate to\\[-12mm]
\beq
z\geq z_c, \qquad \mbox{with}\qquad z_c<\!\!< R.
\eeq  
The AdS-boundary is thus now located at $z=z_c$.  The emission, reflection and reception times of a photon, that is emitted from $z = z_c$ and again reflected from a mirror in the bulk at $z = R$, are now given by
\beq
x_0^{emission}= - R+z_c,
\qquad\quad x_0^{reflection} = 0,
\qquad\quad \mbox{and}\qquad \quad x^{reception}_0=  R-z_c. 
\eeq
Hence, after introducing the cut-off, the round trip time of the photon  equals $2(R-z_c)$: this is equal to twice the difference in the $z$-coordinate of the mirror and the boundary. 

The {\it physical time} that is measured by a clock on the boundary at $z=z_c$ is, however, not equal to $x_0$. Namely, due to the pre-factor $(L/z_c)^2$ in front of the metric, the actual time measured on a physical clock is given by $L x_0/z_c$. The physical round trip time $T_{r.t.}$ of the photon is therefore,
\beq
\label{roundtrip1}
T_{r.t.} = {L\over z_c}\cdot 2(R-z_c). 
\eeq
Note that this round trip time would diverge if we send the cut-off $z_c\to 0$. 

A second motivation for introducing the cut-off $z_c$ is that the  area $A(\Sigma)$ of the Rindler horizon remains finite. Previously we have shown that this area is given by Eqs.~(\ref{areadef}) and (\ref{regularizedvolume}) in terms of another cut-off $\chi_c$.  To determine the relationship between the two cut-offs $z_c$ and $\chi_c$ we observe that the horizon is located at
\beq
z^2+x_i^2=R^2\qquad \mbox{with}\qquad |x_i|= R \tanh \chi. 
\eeq
Imposing that $\chi\leq \chi_c$  leads to the inequalities
\beq
x_i^2 \leq R^2 \tanh^2 \chi_c\qquad\mbox{hence}\qquad z^2\geq R^2 -R^2\tanh^2\chi_c.
\eeq
From these equations we learn that the relationship between $z_c$ and $\chi_c$ is given by
\beq
\label{cutoffs}
z_c= R\sqrt{1-\tanh^2 \chi_c} = {R\over \cosh \chi_c}.
\eeq
In the following we will be interested in comparing the value of the photon round trip time with the horizon area $A(\Sigma)$. For this purpose we express both in terms of the same cut-off parameter $\chi_c$.   After inserting Eq.~(\ref{cutoffs}) into Eq. (\ref{roundtrip1}) we find for the round trip time\\[-6mm]
\beq
\label{roundtrip2}
T_{r.t.} =  2L(\cosh \chi_c-1).\\[-8mm]
\eeq
For later convenience we also repeat here the full expression for the horizon area 
\beq
\label{Asigma}
A(\Sigma) = L^{d-1}\Omega_{d-2} \int_0^{\chi_c} \!\!\! d\chi \,  \sinh^{d-2} \chi.
\eeq
Both these equations hold in arbitrary dimensions. The discussion in
this subsection has been completely classical, and has so 
far ignored any fluctuations in the geometry. The aim of the following subsection is to determine what happens to the location of the horizon and the photon travel times due to the fluctuations in the Newtonian potential $\Phi$. 

\subsection{Fluctuations in horizon location and photon travel time}

By introducing the cut-off $\chi_c$ we have regularized the horizon area $A(\Sigma)$, and turned it into a finite quantity. This implies that the fluctuations in the Newton potential $\Phi$ given in Eq.~(\ref{Phisq}) are non-vanishing. The purpose of this subsection is to discuss the physical implications of the  fluctuations in $\Phi$ on the horizon location and the photon travel time.  
\newcommand{\Phiq}{{\Phi}}

Let us assume that quantum fluctuations cause $\Phi$, at some moment in time, to acquire a small but non-zero value of the order of 
\beq
\label{Phiq}
\Phi \sim  \sqrt{\bigl\langle\Phi^2\bigr\rangle } \ = \ 
{1\over d-1} \sqrt{4G\over A(\Sigma)}. 
\eeq
In other words, we regard ${\Phi}$ as a stochastic variable, whose variance is given by $\langle\Phi^2\rangle$. 
First let us determine the perturbation in the location of the Rindler horizon due to $\Phi$.  In terms of the Rindler-AdS coordinates, it is located at the solution of $f(r_h)=0$. Assuming that  ${\Phi}<\!\!< 1$, we find that the perturbed horizon is now located at 
\beq
\label{newhorizon}
{r_h^2\over L^2}-1 = \Phi.
\eeq
The fluctuations in $\Phi$ thus induce a quantum uncertainty in the location of the horizon. This result is in accordance with the arguments of Ref. \cite{Marolf:2003bb}. It will be useful to rewrite the perturbed location of the horizon given in Eq. (\ref{newhorizon}) in terms of Poincar\'{e} coordinates. Inserting Eq.~(\ref{newhorizon}) into Eq.~(\ref{lightcone}) gives 
\beq
\label{lightconepert}
\left({(R-x_0)^2- z^2-x_i^2\over 2Rz}\right)\left({(R+x_0)^2- z^2-x_i^2\over 2Rz}\right)= 2 \Phi.
\eeq
Our next goal is to compute the effect on the travel time of the photon.
For this purpose we will now make the physical assumption that the uncertainty in the location of the reflection point of the photon is equal to the uncertainty in the location of the bifurcate horizon. We will again choose the reflection moment with the mirror to be at $x_0=0$ and to be located at $x_i=0$. Hence, we are only considering fluctuations in the $z$-coordinate.  
One easily works out that the solution of Eq.~(\ref{lightconepert}) for the coordinate $z$,  when we take $x_0=x_i=0$,  is given by
\beq
{R^2- z^2\over 2Rz}= \pm \sqrt {2 \Phiq} \qquad\implies \qquad z^{reflection} = R\left(\sqrt{1+2\Phiq} \pm \sqrt{2\Phiq} \right).
\eeq
Hence the quantum uncertainty in location of the reflection point is 
$\Delta z^{reflection} = R\sqrt{2\Phiq}$,
where we ignored terms of order $\Phi$.  Note that we could have also phrased this calculation as the reflection point in the bulk remaining fixed, but the size of the causal diamond on the boundary shifting from $R \rightarrow \tilde R = R \left(\sqrt{1+2\Phiq} \pm \sqrt{2\Phiq} \right)$.  Either calculation gives the same time delay.  This is because the physical result can be found equally as a (passive or active) coordinate transformation on the boundary or in the bulk.

As we explained, the time it takes  in terms of $x_0$ for a photon to traverse between the mirror and the boundary is equal to the difference in the $z$-coordinates of the mirror and the boundary. This implies that the quantum uncertainty  in the $z$-coordinate directly translates into a quantum uncertainty  in the emission and reception times. We may thus conclude that 
\beq
\label{Deltasquared}
\left.{\Delta z^2\over z^2}\right|_{{reflection}} = \left.{\Delta x_0^2\over x_0^2}\right|_{{emission}} = \left.{\Delta x_0^2\over x_0^2}\right|_{{reception}} = 2\Phiq. 
\eeq
For the physical round trip time for the photon we reach a similar conclusion. In fact, the ratio of the fluctuations $\Delta T_{r.t.}$ to the travel time $T_{r.t.}$ is independent of the metric pre-factor and hence is given in terms of $\Phi$ by a similar expression as Eq.~(\ref{Deltasquared}). 

The statements made thus far involve a single photon trajectory.  Observables, however, involve interference of light following two different trajectories.  Further, $\langle \Phiq \rangle = 0$, such that one must consider the {\em variance} of Eq.~(\ref{Deltasquared}).  Thus the first contribution to the observable enters as a four-point correlation function of the observable:
\beq
\label{fourpt}
\left\langle \left({\Delta T^2_{r.t.}\over T^2_{r.t.}}\right)^2 \right\rangle = 4 \bigl\langle \Phiq^2 \bigr\rangle.
\eeq
In an experiment measuring the phase of light, ${\Delta T^2_{r.t.}\over T^2_{r.t.}}$ is correlated over two different paths. 
Making use of Eq.~\ref{Phiq}, we thus have
\beq
\label{timefluctuations}
{\Delta T^2_{r.t.}\over T^2_{r.t.}} = {2\over d-1} \sqrt{4G\over A(\Sigma)}.  
\eeq
Thus we see, because the first non-zero contribution to fluctuations in time-of-arrival of a pulse of light enters at the four-point, and because the gravitational potential, $\langle \Phiq^2 \rangle \sim \frac{G^2}{L^2} S$ is enhanced by a factor of the entropy (containing $S \sim 1/G$), the fluctuations in the observable scale with $\sqrt{G}$.  

We now compare this surprising result to that obtained in our previous work Ref.~\cite{Verlinde:2019xfb} for $4d$  Minkowski space. For this purpose we take the dimensions to be $d=3$, so that we are dealing with a  $4d$ AdS spacetime. The expression in Eq.~(\ref{Asigma}) for the area $A(\Sigma)$ is in this case easily evaluated,\\[-10mm]
\beq 
A(\Sigma) \ \, {}^{\strut d=3} \!\!\!\!\!\!\!\!=  \ \,  2\pi L^2 \left(\cosh\chi_c-1\right).
\eeq 
By inserting this into our general result, Eq.~(\ref{timefluctuations}), we find for the fluctuations in the photon travel time\\[-12mm]
\beq
{\Delta T^2_{r.t.}\over T^2_{r.t.}} \ \,{}^{\strut d=3} \!\!\!\!\!\!\!\!= \ \,{l_p\over 2\pi L}\cdot{1\over \sqrt{\cosh\chi_c-1}},
\eeq
where we identified the $4d$ Planck scale with $l_p^2 =8\pi G$. We thus confirm the result of Ref.~\cite{Verlinde:2019xfb}, that the fluctuations in the photon travel times scale like $\sqrt{l_p}$, with a denominator that is determined by the light crossing time.   
The difference in the denominator of this expression, relative to our previous result in flat space, is due the negative curvature of AdS space.  Also note that in Ref.~\cite{Verlinde:2019xfb}, the fluctuations in Eq.~\ref{fourpt} were correlated over two different paths, with a correlation strength that (in flat four-dimensional space) is given by spherical harmonics.

Lastly, we note that the flat space results in Ref.~\cite{Verlinde:2019xfb} can be reconstructed in a Randall-Sundrum II scenario, where the boundary is pulled into the bulk, inducing gravity on the brane.  When $\chi_c \gg 1$, we have
\beq
\frac{\Delta  T_{r.t.}^2}{T_{r.t.}^2} ={2^{d/2}\over d-1} \sqrt{\frac{4 (d-2) G_{bulk}}{\Omega_{d-2} L^{d-1} e^{(d-2)\chi_c}}} = {2^{d/2}\over d-1}\sqrt{\frac{4 G_{bdy}}{\Omega_{d-2} L^{d-2} e^{(d-2)\chi_c}}} \ \, {}^{\strut d=4} \!\!\!\!\!\!\!\! \sim  \ \,  \sqrt{\frac{G_{bdy}}{T_{r.t.}^2}}, 
\eeq
where we have used the fact that the bulk and boundary gravitational constants are related by $G_{bdy} = (d-2)G_{bulk}/L$.  

\section{Discussion}

The main aim of this work was to study vacuum energy fluctuations in AdS space that occur in a finite spacetime region, and to study their implications for the spacetime geometry. In particular we analyzed their effect on the travel times of photons that reflect off a mirror in the bulk. An important step in our argumentation was that, for the problem of interest, one may focus on the causal diamond associated with a single round trip of a photon.   We showed that, while the expectation value in the energy remains zero, there exist local energy fluctuations of the size given by Eq.~(\ref{mainresult}) in any finite region of the AdS-spacetime.
It is important to point out, however, that the relative size of these local energy fluctuations, and their effect on the geometry, goes to zero when we take the size of the region to infinity. We have  shown that for a finite region, the energy fluctuations source metric fluctuations in the bulk, given by Eq.~(\ref{PhitoK}).  We subsequently showed that a light beam reflecting off a mirror in the bulk experiences fluctuations in the time it leaves and arrives from the boundary, summarized in Eq.~(\ref{timefluctuations}).  
We again emphasize that these effects only occur because one is measuring a finite part of the spacetime.

While our results in AdS/CFT are in qualitative agreement with those obtained in Minkowski space in Ref.~\cite{Verlinde:2019xfb}, in the Minkowski case one has less control over the underlying microscopics.  In Ref.~\cite{Verlinde:2019xfb}, we only presented a heuristic argument for the size of the fluctuations given by Eq.~(\ref{mainresult}), where in the $4d$ flat space situation the surface $\Sigma$ has an area $A = 4 \pi R^2$ and divides the inside and outside of the causal diamond with finite radius $R$. In this paper we verified these arguments in the context of AdS/CFT, and provided further evidence for the assumptions that were made for the flat space case. In particular, one would like to understand to what extent flat space holographic assumptions, like the ones presented in this paper, are justified.  

There are many interesting directions to pursue to clarify this question.  First, moving the boundary CFT into the bulk, as in a braneworld scenario, one could study energy and metric fluctuations on the brane with Minkowski metric.  Second, $\langle \Delta K^2 \rangle$ could be computed directly with the quantized Einstein-Hilbert metric.  We expect that such a calculation would yield energy fluctuations proportional to the surface area (dividing the inside and out of the causal region) but quadratically UV divergent, similar to the classic calculation of the entanglement entropy \cite{Srednicki:1993im,Solodukhin:2011gn}.  

Ultimately we are seeking to show that, at least where certain soft properties of the quantized Einstein-Hilbert action are concerned, the holographic principle applies equally well in Minkowski space as in the context of AdS/CFT.  There has already been progress in these directions in the context of soft theorems, see for example \cite{Kapec:2016jld,Kapec:2016aqd,Cheung:2016iub,Ball:2019atb}.  The finite causal diamond and its realization in the experimental set-up of an interferometer provides a novel context in which to pursue these questions further.

\bigskip

{\em Note added:} After completing our work, we became aware of references Refs.~\cite{Perlmutter:2013gua, Nakaguchi:2016zqi, deBoer:2018mzv} which have some overlap with our calculation of $\langle \Delta K^2 \rangle$.  We thank Eric Perlmutter for bringing Ref.~\cite{Perlmutter:2013gua} to our attention, which contains a CFT derivation, and a corresponding bulk computation, of the temperature variation of the $n$-th R\'{e}nyi entropy. With suitable identifications the calculations presented in this work agree with our result Eq.~(\ref{mainresult}). Refs. \cite{Nakaguchi:2016zqi}, and \cite{deBoer:2018mzv} consider the same physical quantity, referred to as the capacity of entanglement, and present results consistent with the ones presented here. 

\mysection{Acknowledgments}  We thank Nima Arkani-Hamed, Tom Banks, Dan Carney, Cliff Cheung, David Gross, Cynthia Keeler, Takemichi Okui and Raman Sundrum for discussions.

\bibliography{QG}

\end{document}